\pdfoutput=1
\documentclass[conference]{IEEEtran}
\IEEEoverridecommandlockouts
\usepackage{cite}
\usepackage{amsmath,amssymb,amsfonts}
\usepackage{algorithmic}
\usepackage{graphicx}
\usepackage{url}
\usepackage{float}
\usepackage{textcomp}
\usepackage{xcolor}
\usepackage{hyperref}
\makeatletter
\newcommand\subparagraph{%
  \@startsection{subparagraph}{5}
  {\parindent}
  {3.25ex \@plus 1ex \@minus .2ex}
  {-1em}
  {\normalfont\normalsize\bfseries}}
\makeatother
\usepackage{titlesec}
\let\subparagraph\relax
\usepackage{array}
\newcolumntype{P}[1]{>{\centering\arraybackslash}p{#1}}
\newcolumntype{M}[1]{>{\centering\arraybackslash}m{#1}}

\def\BibTeX{{\rm B\kern-.05em{\sc i\kern-.025em b}\kern-.08em
    T\kern-.1667em\lower.7ex\hbox{E}\kern-.125emX}}
\begin{document}

\newcommand\copyrighttext{
	\Huge {IEEE Copyright Notice} \\ \\
	\large {Copyright (c) 2020 IEEE \\
		Personal use of this material is permitted. Permission from IEEE must be obtained for all other uses, in any current or future media, including reprinting/republishing this material for advertising or promotional purposes, creating new collective works, for resale or redistribution to servers or lists, or reuse of any copyrighted component of this work in other works.} \\ \\
	
	{\Large Accepted to be published in: 2020 IEEE 3rd 5G World Forum (WF-5G), 10-12 September 2020, Virtual Event} \\ \\ 
}
\twocolumn[
\begin{@twocolumnfalse}
	\copyrighttext
\end{@twocolumnfalse}
]

\makeatletter
\newcommand{\newlineauthors}{%
  \end{@IEEEauthorhalign}\hfill\mbox{}\par
  \mbox{}\hfill\begin{@IEEEauthorhalign}
}
\makeatother
\title{Latency Analysis for IMT-2020 Radio Interface Technology Evaluation
}

\author{\IEEEauthorblockN{A. Phani Kumar Reddy}
\IEEEauthorblockA{\textit{Dept. of Electrical Engineering} \\
\textit{Indian Institute of Technology Kanpur}\\
phani.iitk@gmail.com}
\and
\IEEEauthorblockN{Navin Kumar, Sri Sai Apoorva Tirumalasetty}
\IEEEauthorblockA{\textit{Dept. of ECE} \\
\textit{Amrita School of Engineering, Bengaluru}\\
navin\_kum3@yahoo.com, saiapoorva112@gmail.com}
\and

\newlineauthors
\IEEEauthorblockN{Srinivasan S}
\IEEEauthorblockA{\textit{Nokia Bell Labs}\\
srinivasan.selvaganapathy@nokia.com}
\and
\IEEEauthorblockN{Vinosh Babu James J.}
\IEEEauthorblockA{\textit{Qualcomm India}\\
vinosh@qti.qualcomm.com}
\and
}
\IEEEoverridecommandlockouts

\IEEEpubid{\makebox[\columnwidth]{978-1-7281-7299-6/20/\$31.00 \copyright~2020~IEEE }
\hspace{\columnsep}\makebox[\columnwidth]{ }}
\maketitle

\begin{abstract} The International Telecommunication Union (ITU) is currently deliberating on the finalization of candidate radio interface technologies (RITs) for IMT-2020 (International Mobile Telecommunications) suitability. The candidate technologies are currently being evaluated and after a couple of ITU-Radiocommunication sector (ITU-R)  working party (WP) meetings, they will become official. Although, products based on the candidate technology from 3GPP (5G new radio (NR)) is already commercial in several operator networks, the ITU is yet to officially declare it as IMT-2020 qualified. Along with evaluation of the 3GPP 5G NR specifications, our group has evaluated many other proponent technologies. 3GPP entire specifications were examined and evaluated through simulation using Matlab and using own developed simulator which is based on the Go-language. The simulator can evaluate complete 5G NR performance using the IMT-2020 evaluation framework. In this work, we are presenting latency parameters which has shown some minor differences from the 3GPP report. Especially, for time division duplexing (TDD) mode of operation, the differences are observed. It might be possible that the differences are due to assumptions made outside the scope of the evaluation. However, we considered the worst case parameter. Although, the report is submitted to ITU but it is also important for the research community to understand why the differences and what were the assumptions in scenario for which differences are observed.

\end{abstract}


\section{Introduction}
\par The International Telecommunication Union Radiocommunication sector (ITU-R) working party-5D (WP5D) is almost reaching to consensus on international mobile telecommunication (IMT)-2020 (5G) standard specifications \cite{ITUweb}, and most likely to be finalized by 2020. The requirements for these radio access technologies (RAT) is available in ITU-R report M.2410 \cite{ITUmin}. The developers of radio access technologies such as third generation partnership project (3GPP) worked on development of fifth generation (5G) technologies meeting these requirements. 3GPP have developed RATs new radio (NR),
\\
\\

Long term evolution - machine type communication (LTE-M) and narrow band-Internet of Things (NB-IoT) which together meet all requirements specified (3GPP TR38.913)\cite{3gpp.38.913}. Additionally, few other candidate technologies developed by Korea, China, European Telecommunications Standards Institute-digital enhanced cordless telecommunication (ETSI-DECT) Forum, NuFront have been submitted to ITU-R. These candidate technologies said to have met the minimum technology requirements. 
\par The report ITU-R M.2410 \cite{ITUmin} defines 13 minimum requirements related to the technical performance of the IMT-2020 radio interface(s). Recommendation ITU-R M.2083 \cite{ITUframe} defines eight key ''Capabilities for IMT-2020''. Also, reports ITU-R M.2412 \cite{ITUguide},\cite{3gpp.38.211} defines the detail methodology to be used for evaluating the minimum requirements, including test environments, evaluation configurations and, channel models.
\par The user trends for IMT together with the future role
and market lead to a set of usage scenarios envisioned for both human-centric and machine-centric communication. The usage scenarios identified are enhanced. Although, 3GPP defined specification meet the IMT-2020 requirement, these need to be verified through independent evaluation groups (IEG). Along with 3GPP specifications, our group has evaluated many other proponent technologies. The entire 3GPP specifications were examined and evaluated through simulation using Matlab and using own developed simulator. Both simulator are found to have 100 percent accuracy. Our developed simulator uses Go-language. The simulator can evaluate complete NR performance. 
\par In this work, we are presenting latency parameters which have shown some differences from the 3GPP report. Especially, for time division duplexing (TDD) mode, the differences are observed. It might be possible that the differences are because of assumptions. However, we considered the worst case parameters. We used analytical method for evaluation. In fact, steps comprise; 
\\
\\
\\
\\
 self evaluation, inspection, analytical and experimental verification. Our major work involved in inspection and analytical part, but some of the experiments carried out on industry test bed is also performed. The report has been submitted to ITU. 
At the same time, it is also important for the research community to understand why the differences and what were the assumptions in the scenario for which differences are observed.
\par Contents of the rest of the paper is as follows. Section~\ref{Sec: Minimum Requirements} briefly explains latency and minimum requirements. Evaluation methodology is presented in Section~\ref{Sec: Evaluation Methodology}, while observations are discussed in Section~\ref{Sec: Results}. Section~\ref{Sec: Conclusion} finally, concludes the paper.

\section{Latency and Minimum Requirements}\label{Sec: Minimum Requirements}
\subsection{Latency}
Ultra-reliable and low latency communications (URLLC) is one of the three major use cases defined by 3GPP. There are multiple applications like remote medical surgery, rescue operation which require extremely low latency. 

Latency is defined as the average time between the transmission of packet and the reception of an acknowledgment. Several scenarios require the support of very low latency and very high communications service availability. The overall service latency depends on the delay on the radio interface, transmission within the 5G system, transmission to a server which may be outside 5G system and data processing. Some of these factors depend directly on the 5G system itself. Whereas for others the impact can be reduced by suitable interconnections between the 5G system and services or servers outside of the 5G system. For example, to allow local hosting of the services. The latency performance of a communication system is analyzed for both control plane (CP) and user plane. 

\begin{itemize}
    \item Control Plane Latency: According to 3GPP TR 38.913 \cite{3gpp.38.913}, control plane latency is defined as ''the time to move from a battery efficient state e.g. IDLE  to the start of continuous data transfer e.g. ACTIVE''. Considering agreements made during the study item phase of NR, the control plane latency can be analyzed as the transition time from an inactive state to the time to send the first uplink (UL) packet in the inactive state. This requirement is defined for the purpose of evaluation in the enhanced mobile broad band (eMBB) and ultra reliable low latency communication (URLLC) usage scenarios. If a 5G g node B (gNB) is integrated with a LTE eNB, and the control protocol (i.e. radio resource control (RRC)) is located in the LTE eNB, the control plane latency will be the same as in the LTE case.
\item {User Plane Latency:} User plane latency is defined as ''the time to successfully deliver an application layer packet/message from the radio protocol L2/L3 service data unit (SDU) entering point to the radio protocol L2/L3 SDU entering point via the radio interface in both UL and DL directions, where neither device nor the base station reception is restricted by discontinuous reception  (DRX). In other words, the user plane latency is analyzed as the radio interface latency from the time when transmitter packet data convergence protocol (PDCP) receives an Internet protocol (IP) packet to the time when receiver PDCP successfully receives the IP packet and delivers the packet to the upper layer.

\end{itemize}
\subsection{Minimum Requirements}
The IMT-2020 proposal \cite{ITUweb} defines the minimum latency support for control plane and user plane as specified in Table~\ref{Tbl:IMT-2020 Requirements}.
\begin{table}[tbh]
\begin{center}
\caption{IMT-2020 Latency Requirements \cite{ITUmin}.}
\begin{tabular}{|c|c|c|}
\hline
\textbf{Latency} &  \textbf{eMBB} & \textbf{URLLC}  \\
\hline
\textbf{User Plane Latency} & 4ms & 1ms \\
\hline
\textbf{Control Plane Latency} & 20ms$^{\dagger}$ & 20ms$^{\dagger}$ \\
\hline
\multicolumn{3}{l}{$^{\dagger}$Proponents can consider lower control plane latency, e.g. 10ms}
\end{tabular}
\label{Tbl:IMT-2020 Requirements}
\end{center}
\end{table}
\section{Evaluation Methodology}\label{Sec: Evaluation Methodology}
\par As IMT-2020 being considered from multiple perspectives (users, manufacturers, application developers, network operators, service and content providers and finally, the usage scenarios) which are extensive. Therefore, candidate RITs/SRITs for IMT-2020 must be capable of being applied in a much broader variety of usage scenarios. Also, it should support a much broader range of environments, significantly more diverse service capabilities as well as technology options.
\par ITU-R M.2412 \cite{ITUguide} provides detail guideline of evaluation comprising from inspection to experimental evaluation. Our report is mostly based on analytical analysis and completed using developed simulator supported and verified by Matlab. Supporting technology should provide several elements and their values for both control and user plane latency. Table~\ref{Tbl:CP Analysis Template} and Table~\ref{Tbl:UP Analaysis Template} provides the detail requirements for control plane and user plane respectively. 
\begin{table}[tbh]
\caption{Example of control plane latency analysis template \cite{ITUguide}}
\centering
\begin{tabular}{|c|c|}
\hline
\textbf{Step}&\textbf{Description}\\
\hline
1& Random access procedure \\
\hline
2&UL synchronization\\
\hline
3&Connection establishment + HARQ retransmission  \\
\hline
4&Data bearer establishment + HARQ retransmission \\
\hline
 &Total control plane latency = Sum of 1) to 4) \\
 \hline
\end{tabular}
\label{Tbl:CP Analysis Template}
\end{table}
\begin{table}[tbh]
\caption{Example of user plane latency analysis template \cite{ITUguide}.}
\centering
\begin{tabular}{|c|c|}
\hline
\textbf{Step}&\textbf{Description} \\
\hline
1& UE Processing delay \\
\hline
2&Frame alignment \\
\hline
3&TTI for data packet transmission  \\
\hline
4&HARQ retransmission \\
\hline
5&BS processing delay \\
\hline
 &Total one way user plane latency = Sum of 1) to 5) \\
 \hline
\end{tabular}
\label{Tbl:UP Analaysis Template}
\end{table}
\section{Evaluation Results, Observations and Discussion}\label{Sec: Results}
\par In this section, we provide the details of completed evaluation and our observations related to proponent self evaluation submission and minimum requirement.
\subsection{Control Plane Latency}
\par Figure~\ref{Fig:C-plane procedure} shows the procedure for control plane latency.
\begin{figure}[tbp]
\centering
\centerline{\includegraphics[scale=0.55]{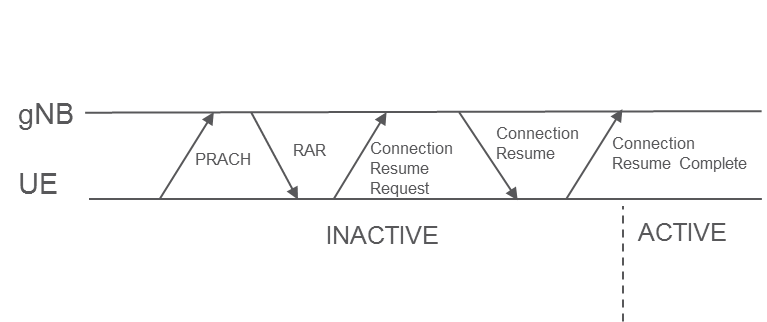}}
\caption{C-plane procedure for evaluation.}
\label{Fig:C-plane procedure}
\end{figure}
\subsubsection{Processing Delay}
We assume that the minimum timing capabilities have been agreed for NR. With the user equipment (UE) capability, the minimum UL timing is set to be 3-symbols for both 15kHz and 30kHz sub-carrier spacing (SCS). For 120kHz, the assumption is made of 9-symbols timing. With mini-slots, the transmit time intervals (TTIs) can have different lengths and therefore, we counted the processing in terms of the shortest considered TTI which is 4-symbols in this paper. For simplicity, the processing delay is set to 1 TTI for both 15 and 30kHz SCS and 3 TTI at 120kHz SCS in both gNB and UE. The RRC processing delays are assumed to be of a fixed value of 3ms \cite{R2}.
\subsubsection{Achievable Latency in frequency division duplexing (FDD)} It is assumed that the UE works with $n$+2 timing and the gNB with $n$+3 timing as the fastest options, i.e. the processing budget is 1 and 2 TTIs respectively. This is for 15 and 30kHz SCS. For 120kHz, the processing delay is doubled in TTIs, giving $n$+3 timing for the UE and $n$+5 timing for gNB.
\par Based on the above assumptions, the resulting CP latency is calculated which is outlined in Table~\ref{Tbl:CP latency in TTIs in NR Rel-15 FDD.}. It can be seen that the total worst-case delay sums up in the range 9-14 TTIs + 6ms for FDD. The worst-case CP latency in NR Rel-15 FDD is estimated to 9 TTI+6ms at 15/30kHz SCS and 14 TTI+6ms for 120kHz SCS.

\begin{table}[htbp]
\caption{CP latency in TTIs in NR Rel-15 FDD.}
\centering
\resizebox{\columnwidth}{!}{\begin{tabular}{|c|c|c|c|}
\hline
\textbf{Component} &\textbf{Description} &\multicolumn{2}{|c|}{\textbf{Latency}}  \\
\hline
& &\textbf{15/30kHz}&\textbf{120kHz} \\
\hline
1& Worst-case delay due to RACH scheduling period&1  TTI
&1 TTI\\
&(1 TTI period)&&\\
\hline
2 & Transmission of \textcolor{red}{RACH Preamble} &1 TTI& 1 TTI\\
\hline
3 &Preamble detection and processing in gNB &1 TTI& 3 TTI\\
\hline
4&Transmission of \textcolor{red}{RA response} &1 TTI &1 TTI\\
\hline
5&UE Processing Delay (decoding of scheduling grant,  &1 TTI& 2 TTI\\
&timing
alignment and C-RNTI assignment &&\\
&+ L1 encoding of RRC
Connection Request&&\\
\hline
6&Transmission of \textcolor{red}{RRC Connection Resume Request} & 1 TTI& 1 TTI\\
\hline
7&Processing delay in gNB & 3 ms& 3 ms\\
&(L2 and RRC)&&\\
\hline
8&Transmission of \textcolor{red}{RRC Connection Resume} & 1 TTI& 1 TTI\\
&(and UL grant)&&\\
\hline
9&Processing delay in the UE & 3 ms &3 ms\\
&(L2 and RRC)&&\\
\hline
10&Transmission of \textcolor{red}{ RRC Connection Resume Complete}  &1 TTI& 1 TTI\\
&(including
NAS Service Request)&&\\
\hline
11&Processing delay in gNB & 1 TTI &3 TTI\\
&($Uu \rightarrow S1-C$)&&\\
\hline
&\textbf{Total delay} &\textbf{9 TTI+6ms}& \textbf{14 TTI+6ms}\\
\hline
\end{tabular}}
\label{Tbl:CP latency in TTIs in NR Rel-15 FDD.}
\end{table}


\subsubsection{Achievable Latency in TDD}
For the TDD slot sequence, two cases are studied: an alternating UL-DL sequence and, a DL- heavy UL-DL-DL-DL sequence. Due to the slot sequence, additional alignment delays are added. With the assumptions described above, the resulting CP latency is given in Table~\ref{Tbl:CP latency in TTIs in NR Rel-15 TDD}. It is seen that the total worst-case delay sums up in the range 12-26 TTI + 6ms for TDD. The worst-case CP latency in NR Rel-15 TDD with alternating UL-DL pattern is estimated to 14 TTI+6ms for 15/30kHz SCS and 20TTI+6ms for 120kHz SCS.
\begin{table*}[htbp]
\caption{CP latency in TTIs in NR Rel-15 TDD.}
\begin{center}
\begin{tabular}{|c|c|c|c|c|c|}
\hline
\textbf{Component} &\textbf{Description} &\multicolumn{2}{|c|}{\textbf{UL-DL Latency}} &\multicolumn{2}{|c|}{\textbf{UL-DL-DL-DL Latency}}  \\
\hline
& &\textbf{15/30kHz}&\textbf{120kHz}&\textbf{15/30kHz}&\textbf{120kHz} \\
\hline
1& Worst-case delay due to RACH scheduling period &
2 TTI& 2 TTI& 4 TTI& 4 TTI\\
\hline

2 &Transmission of \textcolor{red}{RACH Preamble}& 1 TTI &1 TTI &1 TTI& 1 TTI
\\
\hline
3& Preamble detection and processing in gNB &1 TTI &3 TTI &1 TTI& 3 TTI\\
\hline

4& DL slot alignment &1 TTI &1 TTI& 0 TTI &1 TTI\\
\hline

5& Transmission of \textcolor{red}{RA response}& 1 TTI& 1 TTI &1 TTI &1 TTI\\
\hline

6& UE Processing Delay & 1 TTI& 3 TTI &1 TTI &3 TTI\\
\hline

7 &UL slot alignment& 1 TTI &1 TTI &0 TTI& 3 TTI\\
\hline
8 &Transmission of \textcolor{red}{RRC Connection Resume Request}& 1 TTI& 1 TTI &1 TTI &1 TTI\\
\hline
9 &Processing delay in gNB & 3 ms &3 ms& 3 ms &3 ms\\
\hline
10& DL slot alignment &1 TTI &1 TTI &0 TTI& 1 TTI\\
\hline
11 &Transmission of \textcolor{red}{RRC Connection Resume}& 1 TTI& 1 TTI& 1 TTI& 1 TTI\\
\hline
12 &Processing delay in the UE &3 ms &3 ms& 3 ms& 3 ms\\
\hline
13& UL slot alignment &1 TTI &1 TTI &0 TTI &3 TTI\\
\hline
14& Transmission of \textcolor{red}{ RRC Connection Resume Complete} & 1 TTI &1 TTI& 1 TTI &1 TTI\\
\hline
15& Processing delay in gNB  &1 TTI &3 TTI& 1 TTI& 3 TTI\\
\hline
& \textbf{Total delay} &\textbf{14TTI+6ms}&\textbf{20 TTI + 6ms}& \textbf{12 TTI + 6ms} &\textbf{26 TTI + 6ms}\\
\hline
\end{tabular}
\label{Tbl:CP latency in TTIs in NR Rel-15 TDD}
\end{center}
\end{table*}


\subsection{User Plane Latency}
\par Figure~\ref{Fig: UP Latency Procedure} shows the procedure for user plane (UP) latency.
\begin{figure}[htbp]
\centerline{\includegraphics[width=\columnwidth]{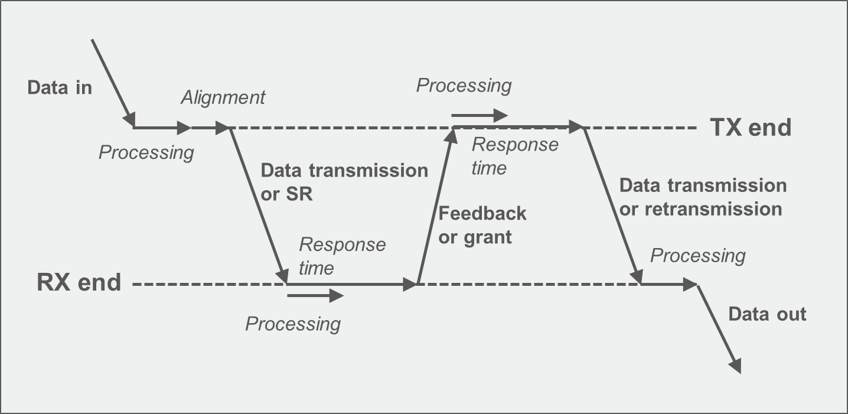}}
\caption{User plane procedure for evaluation.}
\label{Fig: UP Latency Procedure}
\end{figure}

\subsubsection{Processing Delay}
This is the delay caused at the transmitter for preparation of the transmission and at the receiver by reception procedures and decoding.
On the DL, the processing delay in the UE includes the reception and decoding procedure. On the UL, the processing delay in the UE due to the reception and decoding of the UL grant. In gNB processing delay comprises of delay caused by scheduling.
\subsubsection{Alignment delay} The alignment delay is the time required after being ready to transmit until transmission actually starts. The assumption is the worst-case latency, meaning that the alignment delay is assumed to be the longest possible. Physical downlink control channel (PDCCH) and physical uplink control channel (PUCCH) opportunities are assumed to be every scheduled TTI.
\subsubsection{gNB timing} The minimum response time in the gNB between Scheduling Request (SR) and UL grant, and between DL hybrid automatic repeat request (HARQ) and re-transmission is assumed to be 1 TTI. For higher SCS and fewer symbols in the mini-slot, the TTI is shorter and, more TTIs should be used for processing. \par The processing in gNB consists of three main components:
\begin{itemize}
    \item Reception processing (Physical Uplink Shared Channel (PUSCH) processing, SR/HARQ-ACK processing).
    \item Scheduling processing (including SDU/PDU processing for DL).
    \item L1 preparation processing for PDSCH and PDCCH.
\end{itemize}
For simplicity, the gNB processing time is referred to as the total processing time. 
The processing time is a lower limit for gNB response time where the assumptions on gNB processing time are given in Table~\ref{Tbl:gNB Processing Time}.
\par The minimum response timing in the UE between DL data and DL HARQ, and between UL grant and UL data. On the DL, the UE processing time is according to $N_{1}$ value (Table~\ref{Tbl:PDSCH processing time}) for UE capability \#1, while on the UL, the UE processing time is according to $N_{2}$ value (Table~\ref{Tbl:PUSCH preparation procedure time}) for UE capability \#2.
\begin{table}[htbp]
\caption{Processing time (in \# of OFDM symbols) assumptions for gNB.}
\begin{center}
\resizebox{\columnwidth}{!}{\begin{tabular}{|c|c|c|c|c|c|c|}
\hline
\textbf{Timing} &\multicolumn{3}{|c|}{\textbf{15kHz/30kHz SCS}}&\multicolumn{3}{|c|}{\textbf{120kHz SCS}} \\
\hline
\textbf{\# symbols} &\textbf{7os TTI}&\textbf{4os TTI}&\textbf{2os TTI}&\textbf{7os TTI}&\textbf{4os TTI}&\textbf{2os TTI}\\
\hline
\textbf{gNB
processing} & 7& 4& 4& 14& 12&10 \\
\hline
\end{tabular}}
\label{Tbl:gNB Processing Time}
\end{center}
\end{table}

\begin{table}[htbp]
\caption{PDSCH processing time in OFDM symbols for the UE capabilities with
front-loaded DM-RS.}\begin{center}
\begin{tabular}{|c|c|c|c|}
\hline
\textbf{\# Symbols}&\multicolumn{3}{|c|}{\textbf{$N_{1}$} \textbf{PDSCH  (front-loaded DMRS)}} \\
\hline
\textbf{ } & \textbf{15kHz SCS}& \textbf{30kHz SCS}& \textbf{120kHz SCS} \\
\hline
\textbf{Capability 2 } & 3& 4.5& 20\\
\hline
\end{tabular}
\label{Tbl:PDSCH processing time}
\end{center}
\end{table}
\begin{table}[htbp]
\caption{PUSCH preparation procedure time.}
\begin{center}
\begin{tabular}{|c|c|c|c|}
\hline
\textbf{\# Symbols}&\multicolumn{3}{|c|}{\textbf{$N_{2}$}  \textbf{PUSCH  preparation time }} \\
\hline
\textbf{ } & \textbf{15kHz SCS}& \textbf{30kHz SCS}& \textbf{120kHz SCS} \\
\hline
\textbf{Capability 2} & 5&5.5& 36\\
\hline
\end{tabular}
\label{Tbl:PUSCH preparation procedure time}
\end{center}
\end{table}
\par In NR Rel.15 no value (lower than capability 1) for 120kHz SCS was agreed.
\begin{itemize}
    \item $N_{1}$: PDSCH processing time in OFDM symbols for the UE capabilities with front-loaded DM-RS.
    \item $N_{2}$: PUSCH preparation procedure time.
\end{itemize}

\subsubsection{UL scheduling}
For UL data, the scheduling can either be based on SR (Scheduling Request) or SPS (Semi Persistent Scheduling) UL. The assumption is that SR periodicity is 2os (OFDM Symbols) corresponding to the shortest periodicity allowed.
\subsubsection{TTI length and pattern}
We considered a slot lengths of 14-symbols as well as mini-slots of 7, 4, and 2 symbols. For TDD, an alternating DL-UL pattern has been assumed to represent the most latency-optimized setup in a carrier. With TDD, slot/mini-slots of 14, 7, and 4 symbols are used.
\paragraph{FDD} For the case of FDD, the HARQ RTT is $(n+k)$ TTI according to Table~\ref{Tbl:gNB Processing Time} (gNB processing time). The resulting UP latency for SCS of 15, 30 and 120kHz is shown in Table~\ref{Tbl:FDD UP one-way latency}. It can be seen that the 1ms requirement can be reached for SCS 15kHz and up depending on mini-slot configuration. On the UL, configured grants reduce the latency considerably compared to SR-based scheduling.

\begin{table*}[htbp]
\caption{FDD UP one-way latency for data transmission with HARQ-based retransmission.}
\begin{center}
\resizebox{\textwidth}{!}{\begin{tabular}{|c|c|c|c|c|c|c|c|c|c|c|c|c|c|}
\hline
\textbf{Latency(ms)} &\textbf{HARQ} &\multicolumn{4}{|c|}{\textbf{15kHz SCS}}&\multicolumn{4}{|c|}{\textbf{30kHz SCS}}&\multicolumn{4}{|c|}{\textbf{120kHz SCS}} \\
\cline{3-14} 
\textbf{} & \textbf{}&\textbf{14os TTI}&\textbf{7os TTI}&\textbf{4os TTI}&\textbf{2os TTI}&\textbf{14os TTI}&\textbf{7os TTI}&\textbf{4os TTI}&\textbf{2os TTI}&\textbf{14os TTI}&\textbf{7os TTI}&\textbf{4os TTI}&\textbf{2os TTI} \\
\hline
\textbf{DL data}& 1st Tx & \colorbox{orange}{3.2}&\colorbox{orange}{1.7}&\colorbox{orange}{1.3}&\colorbox{green}{0.86}&\colorbox{orange}{1.7}&\colorbox{green}{0.91}&\colorbox{green}{0.7}&\colorbox{green}{0.48}&\colorbox{green}{0.55}&\colorbox{green}{0.43}&\colorbox{green}{0.38} &\colorbox{green}{0.31}\\
\cline{2-14}
 & 1st Re-Tx & 6.2 &\colorbox{orange}{3.2}&\colorbox{orange}{2.6}&\colorbox{orange}{1.7}&\colorbox{orange}{3.1}&\colorbox{orange}{1.6}&\colorbox{orange}{1.3}&\colorbox{green}{0.96}&\colorbox{orange}{1.1}&\colorbox{green}{0.87}&\colorbox{green}{0.76}&\colorbox{green}{0.63}\\
\cline{2-14}
 & 2nd Re-Tx & 9.2 &4.7&\colorbox{orange}{3.6}&\colorbox{orange}{2.6}&\colorbox{orange}{4.7}&\colorbox{orange}{2.4}&\colorbox{orange}{2}&\colorbox{orange}{1.5}&\colorbox{orange}{1.6}&\colorbox{orange}{1.3}&\colorbox{orange}{1.1}&\colorbox{green}{0.96}\\
\cline{2-14}
 & 3rd Re-Tx & 12 &6.2&4.6&\colorbox{orange}{3.4}&6.1&\colorbox{orange}{3.1}&\colorbox{orange}{2.7}&\colorbox{orange}{2}&\colorbox{orange}{2.1}&\colorbox{orange}{1.7}&\colorbox{orange}{1.5}&\colorbox{orange}{1.3}\\
\hline
\textbf{UL data (SR)}& 1st Tx & 5.5&\colorbox{orange}{3}&\colorbox{orange}{2.5}&\colorbox{orange}{1.8}&\colorbox{orange}{2.8}&\colorbox{orange}{1.5}&\colorbox{orange}{1.3}&\colorbox{green}{0.93}&\colorbox{orange}{1.2}&\colorbox{orange}{1.1}&\colorbox{green}{1}&\colorbox{green}{0.89}\\
\cline{2-14}
 & 1st Re-Tx & 9.4&4.9&\colorbox{orange}{3.9}&\colorbox{orange}{2.6}&4.7&\colorbox{orange}{2.4}&\colorbox{orange}{2}&\colorbox{orange}{1.4}&\colorbox{orange}{1.9}&\colorbox{orange}{1.7}&\colorbox{orange}{1.6}&\colorbox{orange}{1.3}\\
\cline{2-14}
 & 2nd Re-Tx & 12&6.4&4.9&\colorbox{orange}{3.5}&6.2&\colorbox{orange}{3.2}&\colorbox{orange}{2.6}&\colorbox{orange}{1.9}&\colorbox{orange}{2.6}&\colorbox{orange}{2.3}&\colorbox{orange}{2.1}&\colorbox{orange}{1.8}\\
\cline{2-14}
 & 3rd Re-Tx & 15&7.9&5.9&4.4&7.7&\colorbox{orange}{3.9}&\colorbox{orange}{3.3}&\colorbox{orange}{2.3}&\colorbox{orange}{3.2}&\colorbox{orange}{2.8}&\colorbox{orange}{2.6}&\colorbox{orange}{2.2}\\
\hline
\textbf{UL data (CG)}& 1st Tx & \colorbox{orange}{3.4}&\colorbox{orange}{1.9}&\colorbox{orange}{1.4}&\colorbox{green}{0.93}&\colorbox{orange}{1.7}&\colorbox{green}{0.95}&\colorbox{green}{0.7}&\colorbox{green}{0.48}&\colorbox{green}{0.7}&\colorbox{green}{0.57}&\colorbox{green}{0.52}&\colorbox{green}{0.45}\\
\cline{2-14}
 & 1st Re-Tx & 6.4&\colorbox{orange}{3.4}&\colorbox{orange}{2.6}&\colorbox{orange}{1.8}&\colorbox{orange}{3.2}&\colorbox{orange}{1.7}&\colorbox{orange}{1.4}&\colorbox{green}{0.93}&\colorbox{orange}{1.3}&\colorbox{orange}{1.1}&\colorbox{orange}{1.1}&\colorbox{green}{0.89}\\
\cline{2-14}
 & 2nd Re-Tx & 9.4&4.9&\colorbox{orange}{3.9}&\colorbox{orange}{2.6}&\colorbox{orange}{4.7}&\colorbox{orange}{2.4}&\colorbox{orange}{2}&\colorbox{orange}{1.4}&\colorbox{orange}{1.9}&\colorbox{orange}{1.7}&\colorbox{orange}{1.6}&\colorbox{orange}{1.3}\\
\cline{2-14}
 & 3rd Re-Tx & 12&6.4&4.9&\colorbox{orange}{3.5}&6.2&\colorbox{orange}{3.2}&\colorbox{orange}{2.6}&\colorbox{orange}{1.9}&\colorbox{orange}{2.6}&\colorbox{orange}{2.3}&\colorbox{orange}{2.1}&\colorbox{orange}{1.8}\\
\hline
\multicolumn{14}{l}{1ms (URLLC - green) and 4ms (eMBB-orange)}
\end{tabular}}
\label{Tbl:FDD UP one-way latency}
\end{center}
\end{table*}
\paragraph{TDD} With TDD, there are additional alignment delays caused by the sequence of DL and UL slots. Depending on when the data arrives in the transmit buffer, the latency may be the same or higher than the FDD latency. For a DL-UL pattern with HARQ RTT of ($n$+4) TTI and higher (Table~\ref{Tbl:gNB Processing Time}), the resulting latency is as indicated in Table~\ref{Tbl:TDD UP one-way latency}. The 4ms target can be reached with a SCS of 15kHz for 7-symbol mini slot, while 30kHz SCS is possible also with slot length transmission. The 1ms target can be reached with 120kHz SCS and mini-slots for DL and UL configured grant transmissions.

\begin{table*}[htbp]
\caption{TDD UP one-way latency for data transmission with alternating DL-UL slot pattern.}
\begin{center}
\begin{tabular}{|c|c|c|c|c|c|c|c|c|c|c|}
\hline
\textbf{Latency(ms)} &\textbf{HARQ} &\multicolumn{3}{|c|}{\textbf{15kHz SCS}}&\multicolumn{3}{|c|}{\textbf{30kHz SCS}}&\multicolumn{3}{|c|}{\textbf{120kHz SCS}} \\
\cline{3-11} 
\textbf{} & \textbf{}&\textbf{14os TTI}&\textbf{7os TTI}&\textbf{4os TTI}&\textbf{14os TTI}&\textbf{7os TTI}&\textbf{4os TTI}&\textbf{14os TTI}&\textbf{7os TTI}&\textbf{4os TTI} \\
\hline
\textbf{DL data}& 1st Tx & 4.2&\colorbox{orange}{2.7}&\colorbox{orange}{2.3}&\colorbox{orange}{2.2}&\colorbox{orange}{1.4}&\colorbox{orange}{1.2}&\colorbox{green}{0.68}&\colorbox{green}{0.55}&\colorbox{green}{0.51}\\
\cline{2-11}
 & 1st Re-Tx & 8.2&4.7&4.3&4.1&\colorbox{orange}{2.4}&\colorbox{orange}{2.2}&\colorbox{orange}{1.4}&\colorbox{orange}{1.1}&\colorbox{orange}{1}\\
\cline{2-11}
 & 2nd Re-Tx & 12&6.7&6.3&6.2&\colorbox{orange}{3.4}&\colorbox{orange}{3.2}&\colorbox{orange}{2.2}&\colorbox{orange}{1.6}&\colorbox{orange}{1.5}\\
\cline{2-11}
 & 3rd Re-Tx & 16&8.7&8.3&8.1&4.4&4.2&\colorbox{orange}{2.9}&\colorbox{orange}{2.1}&\colorbox{orange}{2}\\
\hline
\textbf{UL data (SR)}& 1st Tx & 7.5&4.5&4.1&\colorbox{orange}{3.8}&\colorbox{orange}{2.3}&\colorbox{orange}{2.1}&\colorbox{orange}{1.5}&\colorbox{orange}{1.2}&\colorbox{orange}{1.2}\\
\cline{2-11}
 & 1st Re-Tx & 12&6.9&6.4&6.2&\colorbox{orange}{3.4}&\colorbox{orange}{3.2}&\colorbox{orange}{2.3}&\colorbox{orange}{1.9}&\colorbox{orange}{1.7}\\
\cline{2-11}
 & 2nd Re-Tx & 16&8.9&8.4&8.2&4.5&4.2&\colorbox{orange}{3.1}&\colorbox{orange}{2.5}&\colorbox{orange}{2.2}\\
\cline{2-11}
 & 3rd Re-Tx & 20&11&10&10&5.4&5.2&\colorbox{orange}{3.8}&\colorbox{orange}{3.2}&\colorbox{orange}{2.7}\\
\hline
\textbf{UL data (CG)}& 1st Tx & 4.4&\colorbox{orange}{2.9}&\colorbox{orange}{2.4}&\colorbox{orange}{2.2}&\colorbox{orange}{1.4}&\colorbox{orange}{1.2}&\colorbox{orange}{0.82}&\colorbox{green}{0.7}&\colorbox{green}{0.64}\\
\cline{2-11}
 & 1st Re-Tx & 8.4&4.9&4.4&4.2&\colorbox{orange}{2.5}&\colorbox{orange}{2.2}&\colorbox{orange}{1.6}&\colorbox{orange}{1.3}&\colorbox{orange}{1.2}\\
\cline{2-11}
 & 2nd Re-Tx & 12&6.9&6.4&6.2&\colorbox{orange}{3.4}&\colorbox{orange}{3.2}&\colorbox{orange}{2.3}&\colorbox{orange}{1.9}&\colorbox{orange}{1.7}\\
\cline{2-11}
 & 3rd Re-Tx & 16&8.9&8.4&8.2&4.5&4.2&\colorbox{orange}{3.1}&\colorbox{orange}{2.5}&\colorbox{orange}{2.2}\\
\hline
\multicolumn{11}{l}{1ms (URLLC - green) and 4ms (eMBB-orange)}
\end{tabular}
\label{Tbl:TDD UP one-way latency}
\end{center}
\end{table*}
\subsection{Observations}
\subsubsection{Control Plane Latency}
\paragraph{Latency in ms (FDD)} It can be noted that by using SCS of 120kHz the NR can have control plane latency $<$ 10ms. And also, for typical SCS of 15/30kHz the control plane latency is $<$ 20ms.
\begin{table}[htbp]
\caption{Achievable CP latency for NR Rel-15 in ms for FDD.}
\begin{center}
\resizebox{\columnwidth}{!}{\begin{tabular}{|c|c|c|c|}
\hline
\textbf{CP Latency(ms)} &  \textbf{15kHz SCS} & \textbf{30kHz SCS}&\textbf{120kHz SCS}  \\
\hline
\textbf{14-symbol TTI} & 15 (TTI=1ms) & 10.5&7.8 \\
\hline
\textbf{7-symbol TTI} & 10.5 (TTI=0.5ms) & 8.3&6.9 \\
\hline
\textbf{4-symbol TTI} & 8.6 (TTI=0.2888ms) & 7.3&6.5 \\
\hline
\end{tabular}}
\label{Tbl:Achievable CP latency}
\end{center}
\end{table}

\begin{table}[htbp]
\caption{Achievable CP latency for NR Rel-15 in ms for TDD with alternating UL-DL pattern.}
\begin{center}
\begin{tabular}{|c|c|c|c|}
\hline
\textbf{CP Latency(ms)} &  \textbf{15kHz SCS} & \textbf{30kHz SCS}&\textbf{120kHz SCS}  \\
\hline
\textbf{14-symbol TTI} & 20 & 13&8.5 \\
\hline
\textbf{7-symbol TTI} & 13 & 9.5&7.3 \\
\hline
\textbf{4-symbol TTI} & 10 & 8.0&6.7 \\
\hline
\end{tabular}
\label{Tbl:Achievable CP latency-TDD with alternating UL-DL pattern.}
\end{center}
\end{table}
\begin{table}[htbp]
\caption{Achievable CP latency for NR Rel-15 in ms for TDD with UL-DL-DL-DL pattern.}
\begin{center}
\begin{tabular}{|c|c|c|c|}
\hline
\textbf{CP Latency(ms)} &  \textbf{15kHz SCS} & \textbf{30kHz SCS}&\textbf{120kHz SCS}  \\
\hline
\textbf{14-symbol TTI} & 18 & 12&9.3 \\
\hline
\textbf{7-symbol TTI} & 12& 9.0&7.6 \\
\hline
\textbf{4-symbol TTI} & 9.4 &7.7&6.9 \\
\hline
\end{tabular}
\label{Tbl:Achievable CP latency-TDD with alternating UL-DL-DL-DL pattern.}
\end{center}
\end{table}
\paragraph{Latency in ms (TDD)} With different TTI lengths and SCSs, the absolute delay will differ, as shown in Table~\ref{Tbl:Achievable CP latency}. 
From the Table~\ref{Tbl:Achievable CP latency}, all considered configurations fulfil the 20ms 5G target on CP latency and, almost all configurations also reach the 10ms target.
\par Similarly, from the Tables~\ref{Tbl:Achievable CP latency-TDD with alternating UL-DL pattern.} \& Table~\ref{Tbl:Achievable CP latency-TDD with alternating UL-DL-DL-DL pattern.}, all considered configurations fulfil the 20ms 5G target on CP latency for the alternating UL-DL TDD pattern, and several configurations can also fulfill the 10ms requirement.
\par From the assessment of \textbf{Control Plane Latency} following points can be observed.
\begin{enumerate}
    \item The worst-case CP latency in NR Rel-15 FDD is estimated to 9 TTI+6ms at 15/30kHz SCS and 14 TTI+6ms at 120kHz SCS.
    \item NR Rel-15 FDD can reach the 3GPP and ITU 5G targets on CP latency.
    \item The worst-case CP latency in NR Rel-15 TDD with alternating UL-DL pattern is estimated to 14 TTI+6ms for 15/30kHz SCS and 20TTI+6ms for 120kHz SCS.
    \item NR Rel-15 TDD can reach the ITU and 3GPP 5G targets on CP latency. The same is summarized in Table~\ref{Tbl: CP Observations}.
\end{enumerate}
\begin{table}[hbt!]
\caption{Control Plane Latency Observations.}
 \centering
 \resizebox{\columnwidth}{!}{\begin{tabular}{|M{2.5cm}|M{1cm}|M{1.5cm}|M{1cm}|M{1.5cm}|M{2.5cm}|}
  \hline
  \textbf{Minimum technical performance requirements item} &  \textbf{Category} & \textbf{Required  value}&\textbf{Obtained Value} &\textbf{Requirement met?} &\textbf{Comment} \\
  \hline
   \textbf{Control plane latency (ms)} & eMBB & 20 & 8.5-20 & Yes & Various TTI duration, flexible UL \& DL format \\
  \cline{2-5}
  & URLLC & 20 & 6.5-10 & Yes & and SCS allows  to achieve CP latency below 20ms in both FDD \& TDD \\
  \hline
 \end{tabular}}
\label{Tbl: CP Observations}
\end{table}
\subsubsection{User Plane Latency}
\par From the assessment of \textbf{User Plane Latency} following points can be observed.
\paragraph{eMBB}
\begin{itemize}
    \item Can meet both 4ms UP latency on DL even with SCS=15kHz.
    \item Can meet the 4ms UP latency on UL with Scheduled Request at SCS=15kHz, but 1ms UP latency are achievable in limited configurations.
\end{itemize}
\paragraph{URLLC}
\begin{itemize}
    \item Can meet the 1ms UP latency on DL using \textit{\textbf{mini-slots}} at SCS=15kHz.
    \item Can meet 1ms UP latency on UL using ''\textit{\textbf{Configured Grants}}'' at SCS=15kHz and mini-slots.
\end{itemize}
\par User Plane Latency observations shown in Table~\ref{Tbl: UP Observations}.
\begin{table}[hbt!]
\caption{User Plane Latency Observations.}
 \centering
 \resizebox{\columnwidth}{!}{\begin{tabular}{|M{2.5cm}|M{1cm}|M{1.5cm}|M{1.25cm}|M{1.5cm}|M{2.5cm}|}
  \hline
  \textbf{Minimum technical performance requirements item} &  \textbf{Category} & \textbf{Required  value}&\textbf{Obtained Value} &\textbf{Requirement met?} &\textbf{Comment} \\
  \hline
  \textbf{User Plane Latency (ms)} & eMBB & 4 & 0.86-3.9 & Yes &Using various TTI duration (mini-slots), flexible UL \& DL format \\
  \cline{2-5}
  & URLLC & 1 & 0.31-0.96 & Yes & and SCS allows  to achieve required UP latency in both FDD \& TDD \\
  \hline
 \end{tabular}}
\label{Tbl: UP Observations}
\end{table}

\section{Conclusion}\label{Sec: Conclusion}
\par In this paper, we worked out on the latency analysis of IMT-2020 radio interface technology. For all conditions, CP and UP latency is calculated. We conclude that the CP and UP latency for 5G NR are in compliance with IMT-2020 requirements for both FDD and TDD modes. Based on these evaluations, we were able to recommend the acceptance of the 3GPP 5G NR technology as a valid IMT-2020 technology.

\section*{Acknowledgment}

We would like to thank the COAI-5GIF (Cellular Operators Association of India-5G India Forum), and the industry mentors for giving us an opportunity to participate in the 5GIF IMT-2020 independent evaluation group activity, and evaluate the 3GPP 5G-NR candidate technology.
\bibliographystyle{IEEEtran}
\bibliography{main}
\end{document}